\documentclass[a4paper, 12pt, journal, web]{ieeecolor}

\usepackage{tmi}
\usepackage{cite}
\usepackage{amsmath,amssymb,amsfonts}
\usepackage{algorithmic}
\usepackage{graphicx}
\usepackage{subcaption} 
\usepackage{hyperref}
\hypersetup{
    colorlinks=true,
    linkcolor=blue,
    filecolor=magenta,      
    urlcolor=cyan,
}
\usepackage{textcomp}
\def\BibTeX{{\rm B\kern-.05em{\sc i\kern-.025em b}\kern-.08em
    T\kern-.1667em\lower.7ex\hbox{E}\kern-.125emX}}
\begin{document}
\title{Conditional Deep Convolutional Neural Networks for Improving the Automated Screening of Histopathological Images}
\author{Gianluca Gerard, \IEEEmembership{Member, IEEE}, Marco Piastra
\thanks{Submitted for review May 26, 2021.}
\thanks{G. Gerard was with the Department of Electrical, Computer and Biomedical Engineering,
University of Pavia, Pavia, 27100 Italy. He is
now with Sorint.Tek, Grassobbio, BG 24050 Italy (e-mail: gianluca.gerard@latek.it).}
\thanks{M. Piastra is with
the Department of Electrical, Computer and Biomedical Engineering,
University of Pavia, Pavia, 27100 Italy (e-mail: marco.piastra@unipv.it).}
}

\maketitle


\begin{abstract}

Semantic segmentation  of  breast  cancer metastases in histopathological slides is a challenging task.
In fact, significant variation in data characteristics of histopathology images (domain shift) make generalization of deep learning to unseen data difficult. Our goal  is  to  address  this  challenge  by  using a  conditional Fully Convolutional Network (co-FCN) whose output can be conditioned at run  time, and which can improve its performance when a properly selected set of reference slides are used to condition the output.
We adapted to our task a co-FCN originally applied to organs segmentation in volumetric medical images
and we trained it on the Whole Slide Images (WSIs) from three out of five medical centers present in the CAMELYON17 dataset.
We tested the performance of the network  on  the  WSIs  of  the  remaining  centers.
We also developed an automated selection strategy for selecting  the  conditioning  subset, based on an unsupervised clustering process applied to a target-specific set of reference patches,
followed by a selection policy that relies on the cluster similarities with the input patch.
We benchmarked  our  proposed  method  against  a  U-Net  trained  on  the  same  dataset with  no  conditioning. 
The conditioned network shows better performance  that  the  U-Net  on  the  WSIs  with  Isolated Tumor
Cells and micro-metastases from the medical centers used as test. Our contributions are an architecture
which can be applied to the histopathology domain and an automated procedure for the selection of
conditioning data.
\end{abstract}


\section{Introduction}
The prognosis of breast cancer depends on whether the cancer has spread to other organs \cite{Voogd2001}.
For diagnostic purposes, the lymph-nodes closest to the tumor are first removed via biopsy and then screened to detect the presence or absence of metastasizing cancer cells \cite{CRONIN2018320}.
Tissue samples, after fixation and wax embedding, are cut in thin slices, then stained and transferred on glass slides for examination under light microscope.
This visual examination process is labor intensive and time consuming, but the advent of digital pathology with the digitization of glass slides together with the progresses in deep neural networks have opened up the prospect of partially automating the entire screening process \cite{Litjens2016}. 
Notable progresses have been made in the field \cite{Dimitriou2019, Wang2019, Srinidhi2021}. However applications in everyday clinical practice are still very limited.
Among the roadblocks to wide acceptance, remarkable issues are about the robustness and reliability of deep learning (DL) models, which in histopathology are affected by the problem of "domain shift" \cite{Stacke2019}.
This occurs when variations in the acquisition and processing pipeline of slides, variations in the patient population as well as variability among pathologists diagnostic habits contribute to domain variations. These variations can negatively affect the performance of a trained DL model when applied to a dataset distribution which is even slightly different from the one being used for training.
With classical fully-supervised training, a brute-force approach to mitigate domain shift would be to acquire larger training datasets produced through different pipelines, possibly in different centers, to improve the robustness and generalization capabilities of the model.
However obtaining large amounts of high-quality annotated data, showing enough diversity, is very expensive (see \cite{Srinidhi2021}).
A further limiting factor for acceptance of DL in clinical pipelines is model interpretability \cite{Komura2018}: DL algorithms are seen as "black-boxes" \cite{Srinidhi2021}. For a model to be accepted in the clinical practice, it must provide information for its decisions to be vetted.

Our work aims at addressing the challenges above by making the following contributions:
\begin{itemize}
    \item we address the need to limit the extent and cost of acquiring an extensive training dataset by applying Few-Shot Learning (FSL) to histopathology. With FSL, DL algorithms can be trained with a limited set of training examples while maintaining good generalization capabilities;
    \item we adapt a co-FCN, whose pixel-wise predictions can be inspected by pathologists, to lesion segmentation in histopathology slides;
    the additional visual and spatial cues provided by the semantic segmentation provides insights into the underlying interpretation of the network;
    \item we devised a process for the selection of the input used by the network at inference time to adapt and improve the final output results.
\end{itemize}

The results we present do not rely on transfer learning \cite{Raghu2019}, stain normalization \cite{Stacke2019} or other techniques used by others to address domain variations \cite{Ren2019}.
For benchmark, we compare our results against a U-net \cite{Ronneberger2015} architecture trained and tested on the same dataset.

\section{Related work}

Early work of applying FSL for segmentation tasks is described in \cite{Rakelly2018}. The proposed architecture has two branches:
\begin{itemize}
    \item one conditioning branch extracts a latent task representation from a training set;
    \item the segmenter branch segments a test image guided by the latent task representation.
\end{itemize}
This architecture can be trained with annotations being either full or sparse, i.e. where only a few pixels of the training set are annotated, and it was successfully applied to the segmentation of natural images.

\cite{GuhaRoy2020} extends the work of \cite{Rakelly2018} by introducing strong interactions at multiple locations between the conditioning and segmenter branches, instead of only one interaction. This FSS was trained from scratch, without requiring a pre-trained model, and it was successfully applied to the segmentation of volumetric medical images.

\cite{Medela2019} proposes a Siamese Neural Network \cite{VanderSpoel2015} over a dataset of colon tissue images, and applies it as a feature extractor for a dataset composed by healthy and tumoral samples of colon, breast and lung tissue. The resulting representations of the images is used to train a classifier that can perform the classification task with few samples.

\cite{Wang2019} introduces an architecture with two separate arms for the detection and segmentation tasks, respectively. Only a portion of the dataset is densely annotated with segmentation label and the rest is weakly labeled with bounding boxes.  The model is trained jointly in a multi-task learning setting.

\cite{Zhao2019} presents an automated data augmentation method for synthesizing labeled medical images. The authors demonstrate the method on the task of segmenting MRI brain scans. The method requires only a single segmented scan, and leverages other unlabeled scans in a semi-supervised approach.

\cite{Feyjie2020} proposes a model-agnostic FSL framework for semantic segmentation.
The model is trained on episodes, which represent different segmentation problems, each of them trained with a small labeled dataset. Unlabeled images are also made available at each episode. They include surrogate tasks for semantic feature learning.

\section{Methods and dataset}

\subsection{Network architecture}
\label{ssec:arch}

\begin{figure}
    \centering
    \includegraphics[width=0.5\textwidth]{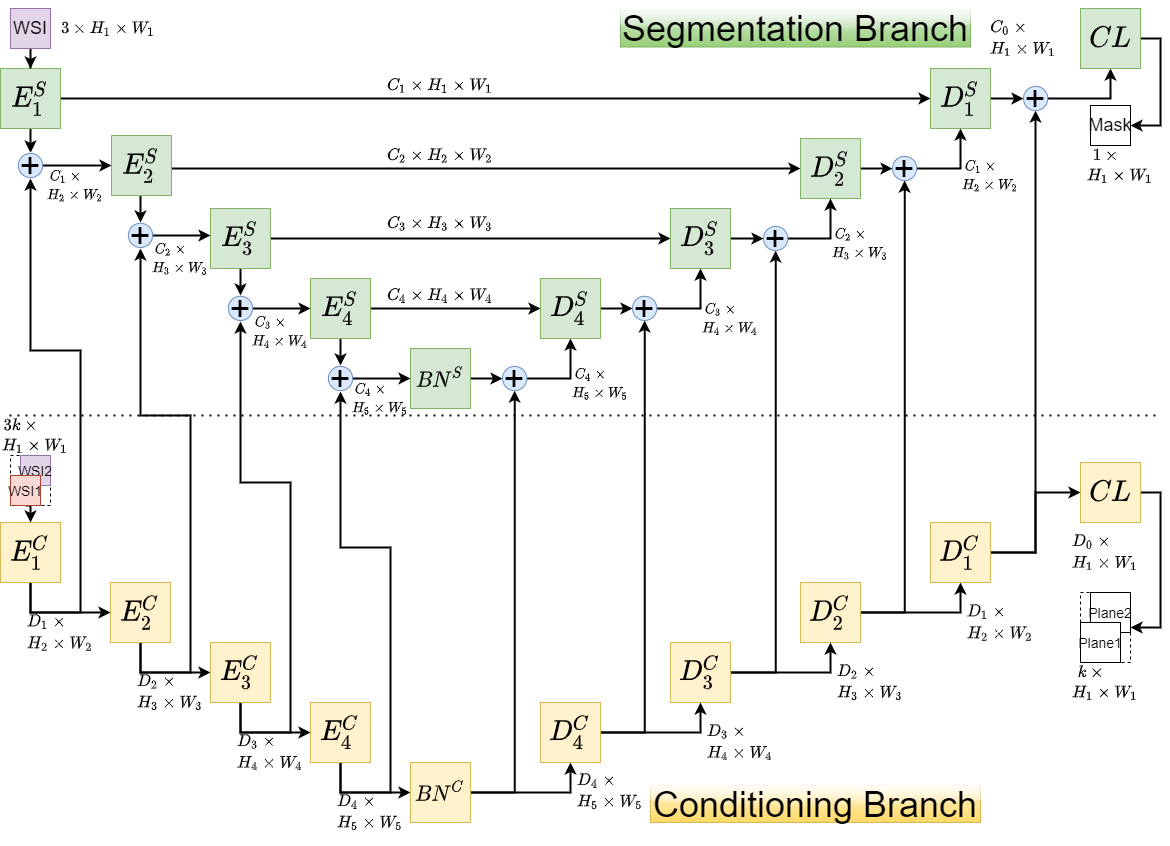}
    \caption{The two branch network used for the co-FCN. $E_i^B$ and $D_i^B$ are the  encoder and decoder blocks in the $B$ branch, which is either Segmentation ($S$) or the Conditioning ($C$).}
    \label{fig:two-branch-arch}
\end{figure}

\begin{figure}
    \centering
    \includegraphics[width=0.35\textwidth]{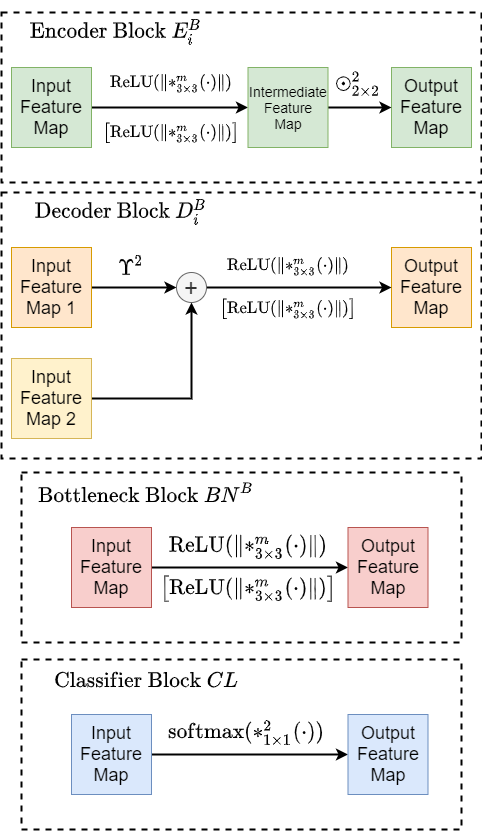}
    \caption{The blocks used in the co-FCN architecture. Connections are explained in Table \ref{tab:connections}.
    }
    \label{fig:ed_blocks}
\end{figure}

Our network architecture is inspired by the FSS described in \cite{GuhaRoy2020}. The network shown in Fig. \ref{fig:two-branch-arch} has two branches:
the Segmentation ($S$) and the Conditioning ($C$) Branch are a sequence of encoder blocks $E_i^B$
followed by a sequence of decoder blocks $D_i^B$.
Each branch is similar to a U-Net \cite{Ronneberger2015}.
The block types used by the network are shown in Fig. \ref{fig:ed_blocks} together with the bottleneck $\textit{BN}^B$ and classifier $\textit{CL}$ blocks. The meaning of the connections are described in Table \ref{tab:connections}.
\begin{table}
    \caption{Operators and functions used in the network blocks in Fig. \ref{fig:ed_blocks}.}
    \label{tab:connections}
    \setlength{\tabcolsep}{3pt}
    \begin{tabular}{p{50pt}|p{180pt}}
    \hline
    \textbf{Symbol} & \textbf{Description} \\
    \hline
         $*^{m}_{p \times q}$ & Convolution: $m$ channels, kernel $p \times q$, stride 1\\
             \hline
         $\mathrm{ReLU}(\cdot)$, $\mathrm{softmax}(\cdot)$, $\sigma(\cdot)$ & ReLU, SoftMax and sigma functions\\
             \hline
         $\left \lVert \cdot \right \rVert$ & Batch normalization\\
             \hline
         ${\odot}^s_{p \times q}$ & Maxpooling with kernel size $p \times q$ and stride $s$\\
             \hline
         $\Upsilon^2$ & Bilinear upsampling with scale factor 2\\
    \hline 
    \end{tabular}
\end{table}
The two branches are connected via channel-wise feature concatenations, which replace the `spatial SE' blocks ($SE$) of the reference architecture in \cite{GuhaRoy2020}. In passing, our choice is similar to how these two branches are connected in the architecture of \cite{Rakelly2018}.
In Fig. \ref{fig:two-branch-arch} the dimensions of the feature maps are shown next to the blocks connections: the actual sizes are described in Table \ref{tab:conn_shapes}.
\begin{table}
    \caption{The dimensions of the features maps flowing through the connections in Fig. \ref{fig:two-branch-arch}.}
    \label{tab:conn_shapes}
    \setlength{\tabcolsep}{3pt}
    \centering
    \begin{tabular}{p{60pt}|p{60pt}}
        \hline
        \textbf{Parameter} & \textbf{Value} \\
        \hline
         $H1$, $W1$ & 128 \\
         $H2$, $W2$ & 64 \\
         $H3$, $W3$ & 32 \\
         $H4$, $W4$ & 16 \\
         $H5$, $W5$ & 8 \\
         $C1$, $D1$, $D2$ & 32 \\
         $C2$, $D3$ & 64 \\
         $C3$, $D4$ & 128 \\
         $C4$ & 256 \\
         \hline
    \end{tabular}
\end{table}

The Segmentation Branch predicts segmentation masks of the input WSI patches. The Conditioning Branch conditions the Segmentation Branch predictions using $k$ WSI patches selected according to the policy described in Subsection \ref{ssec:sset}. Altogether, we refer to the network architecture in \ref{fig:two-branch-arch} as \emph{conditional FCN} (co-FCN).
The Conditioning Branch also outputs, through its classifier $\textit{CL}$, a feature map of dimension $k \times H_1 \times W_1$. During training, this output is averaged across all dimensions and passed through a sigmoid function to predict the probability that the input patches are extracted from lesion regions.
During training and for regularization, the Conditioning Branch classifier's output is compared against the estimated probability derived from the data (see Eq. \ref{eq:prevalence}).
In the Segmentation Branch, we use skip connections to enhance the spatial resolution of the segmentation output.

Skip connections were explicitly removed by \cite{GuhaRoy2020} in their architecture as they observed that caused "the network [to copy] the binary mask of the support set to the output". It has to be remarked that, in our approach, the conditioning input does not contain the binary segmentation masks; this is in contrast to the general approach of providing the conditioning images together with their corresponding dense or sparse annotations (see \cite{Rakelly2018, GuhaRoy2020}). 

\subsection{Dataset}
\label{ssec:dataset}

For training and test we use the CAMELYON17 \cite{Litjens2018} dataset. This dataset contains lesion-level annotations for 50 slides acquired by 5 medical centers.
Table \ref{tab:cam_centers} maps each scanner model to the medical center ID and medical center acronyms in CAMELYON17 as defined in \cite{Litjens2018}.
\begin{table}
\caption{The association between digital scanner and medical centers in the CAMELYON17 dataset.}
\label{tab:cam_centers}
\setlength{\tabcolsep}{3pt}
\begin{tabular}{p{150pt}|p{34pt}|p{44pt}}
\hline
Scanner Model&
Centers ID&
Centers Acronym \\
\hline
3DHistech Panoramic Flash II 250 & 0, 1, 3 & CWZ, RST, RUMC \\
Hamamatsu NanoZoomer-XR C12000-0  & 2 & UMCU \\
Philips Ultrafast Scanner & 4 & LPON \\
\hline
\end{tabular}
\end{table}

To mimic a possible clinical use scenario, in which the network would be trained on a dataset of WSIs acquired by medical centers different from the ones that could actually be using the network, we partitioned the WSIs among training and test sets according to the following rules:
\begin{itemize}
    \item the test and training set WSIs must be acquired by separate medical centers;
    \item a portion of WSIs in the test set must be acquired with a digital scanner different from the ones used to acquire the WSIs of the training set.
\end{itemize}
Therefore, we trained on WSIs from medical centers 0, 1 and 2 and we tested on the WSIs from centers 3 and 4.

For FSL, WSIs must be further divided in two subsets: \emph{support} and \emph{query} (see \cite{Snell2017}). The WSIs in the support are the input to the Conditioning Branch, while the WSIs in the query set are the input to the Segmentation Branch. In distributing the WSIs between these sets we wanted to avoid unbalances between the specific features of the slides. Therefore we stratified the distribution of WSIs based on the classification of patients in CAMELYON17.
This classification is based on the number and type of metastases: macro-metastases, metastases greater than 2 mm; micro-metastases, metastases smaller than 2 mm and greater than 0.2 mm or more than 200 cells; Isolated Tumor Cells (ITCs), single tumor cells or cluster of tumor cells smaller than 0.2 mm or less than 200 cells. Each patient in CAMELYON17 is assigned a \emph{pN-stage} category based on the number and type of metastases: \emph{pN0}, no metastases; \emph{pN0(i+)}, only ITCs; \emph{pN1mi}, micro-metastases present, but no macro-metastases; \emph{pN1}, metastases in 1-3 lymph nodes, at least one a macro-metastasis; \emph{pN2}, metastases in 4-9 lymph nodes, at least one a macro-metastasis.
The selected distribution of patients, manually balanced  between support and query sets by pN-stage and by medical center, is shown in Table \ref{tab:cam_stage}.
\begin{table}
    \caption{Number of patients, per center and per stage classification, in the query set (first number of each column) and in the support set (second number).}
    \label{tab:cam_stage}
    \setlength{\tabcolsep}{3pt}
    \begin{tabular}{p{44pt}|p{34pt}|p{34pt}|p{34pt}|p{34pt}|p{34pt}}
        \hline
        Center ID &
        pN0 &
        pN0(i+) &
        pN1 &
        pN1mi &
        pN2 \\
        \hline
        0 & 0, 0 & 1, 0 & 3, 1 & 2, 0 & 0, 0 \\
        1 & 0, 0 & 1, 0 & 2, 1 & 1, 2 & 1, 0 \\
        2 & 0, 0 & 2, 0 & 2, 0 & 1, 1 & 2, 1 \\
        3 & 0, 0 & 0, 2 & 2, 2 & 1, 2 & 0, 1 \\
        4 & 0, 1 & 1, 1 & 1, 1 & 1, 1 & 0, 2 \\
        \hline
    \end{tabular}
\end{table}
We also balanced the distribution among support and query sets by the type of lesions each WSI contained (see Table \ref{tab:cam_classes}).
\begin{table}
    \caption{Number of slides, per center and per slide classification, in the query set (first number of each column) and in the support set (second number).}
    \label{tab:cam_classes}
    \setlength{\tabcolsep}{3pt}
    \begin{tabular}{p{44pt}|p{34pt}|p{34pt}|p{34pt}|p{34pt}}
        \hline
        Center ID &
        ITC & 
        macro &
        micro & 
        negative \\
        \hline
        0 & 2, 1 & 3, 1 & 2, 1 & 0, 0 \\
        1 & 2, 0 & 1, 1 & 3, 1 & 1, 1 \\
        2 & 2, 0 & 3, 1 & 3, 1 & 0, 0 \\
        3 & 1, 3 & 1, 2 & 1, 2 & 0, 0 \\
        4 & 1, 3 & 1, 2 & 1, 0 & 0, 2 \\
        \hline
    \end{tabular}
\end{table}

We used slides at 20x magnification ($0.5 \mu\mathrm{m} \; \mathrm{pixel}^{-1}$).
For ease of processing, we split the WSIs in a grid of patches, each 128 x 128 pixels wide. 
We discarded patches containing no tissue: patches whose maximum intensity value of the saturation channel, after having applied a Gaussian blur, resulted below 10\% of full saturation.
We retained all patches that had at least 50\% of the pixels in the 64x64 central region classified as lesion.
Of the remaining patches, due to heavy class imbalance, we dropped 85\% of them in the support set and 95\% in the query set.

\subsection{Selection of the support set}
\label{ssec:sset}

As explained, the co-FCN expects two inputs: one patch for the Segmentation Branch (the query patch) and one or more patches from the support set (the support patches) for the Conditioning Branch\footnote{In FSL literature, patches extracted from the support set also known as support \emph{shots}.}.

Our initial choice was to pass a random selection of patches as support, extracted from a pool of lesion and non-lesion patches, one shot each. This is analogous to the policy adopted, for natural images, by \cite{Rakelly2018}, which aimed at training the network to identify similarities and differences between the query patches and the $k$ support shots. However, when the network is presented a random stream of (query, support) pairs, the network does not learn to extract the information from the support necessary to condition the segmentation of the query patches.

Having ascertained the latter experimentally, we opted for a different association policy. To associate each query patch to its corresponding support set we followed the three steps procedure described below.

\subsubsection{Latent representation of the patches with a convolutional autoencoder}
\label{sssec:autoencoder}
\begin{figure}
\centering
\begin{subfigure}{.4\textwidth}
  \includegraphics[width=\textwidth]{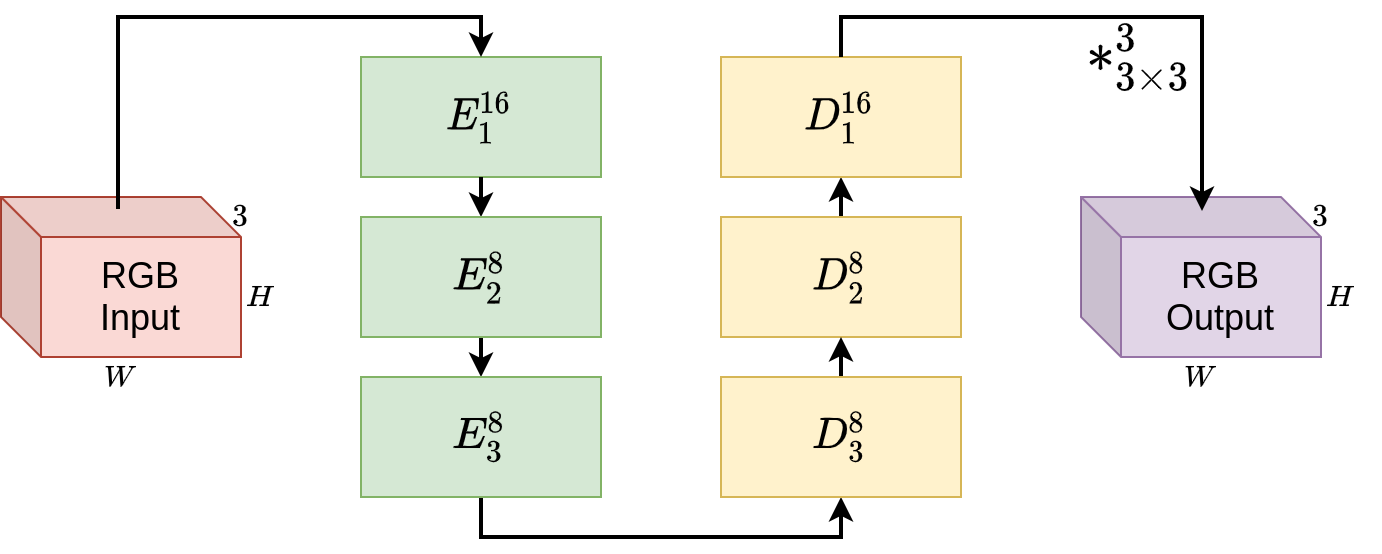}
  \caption{Convolutional autoencoder architecture}
  \label{fig:ae}
\end{subfigure}
\hspace{0.1\textwidth}
\begin{subfigure}{.3\textwidth}
  \includegraphics[width=\textwidth]{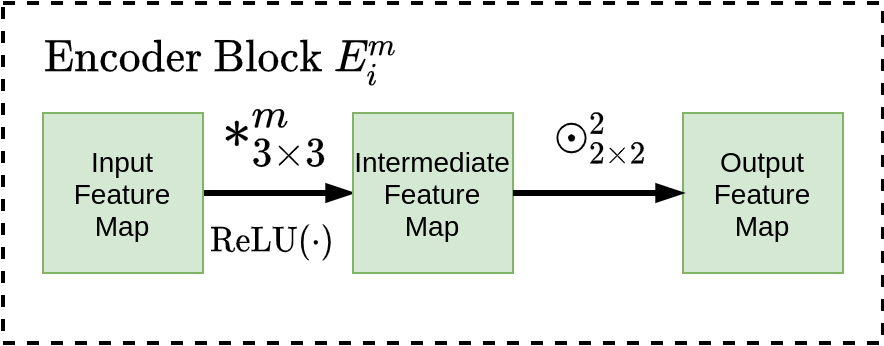}
  \caption{Encoder Block $E_i^m$}
  \label{fig:ae_encoder}
\end{subfigure}
\hspace{0.1\textwidth}
\begin{subfigure}{.3\textwidth}
  \includegraphics[width=\textwidth]{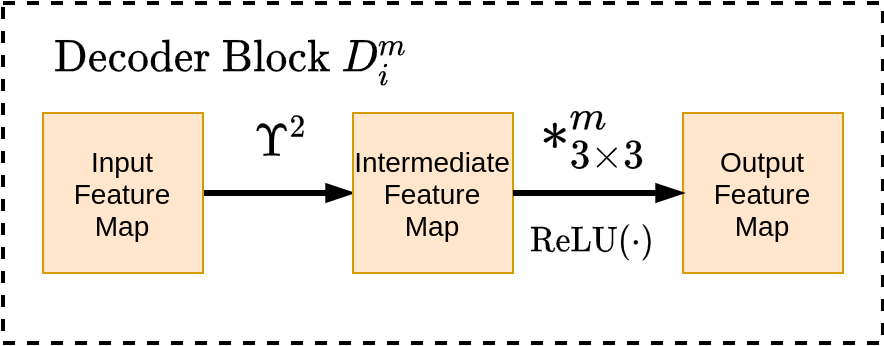}
  \caption{Decoder Block $D_j^m$}
  \label{fig:ae_decoder}
\end{subfigure}
\caption{The autoencoder used for unsupervised learning, with its building blocks.}
\label{fig:autoencoder}
\end{figure}

We trained a convolutional autoencoder (see Fig. \ref{fig:autoencoder}) for each medical center on 80\% of its support patches.
Each autoencoder is trained with an Adam optimzer (with initial learning rate $0.004$, $\beta_1$ 0.9, and $\beta_2$ 0.999). Early stopping was used to avoid overfitting.
The initial weights of each autoencoder were set by training an identical autoencoder on a sample of WSIs from the CAMELYON16 dataset.
The encoder part of the autoencoder extracts a latent representation for each support patch.

\subsubsection{Unsupervised clustering of the support patches from each medical center}
\label{ssec:clustering}
the latent representation was spatially averaged to produce 8 dimensional features vectors, which were reduced via Principal Component Analysis (PCA) to three dimensional vectors. We clustered these vectors with Gaussian Mixture Models (GMMs) applied separately to each medical center (for all medical centers we chose to cluster the vectors in 6 clusters).
In Table \ref{tab:gmm_results} we show the percentage of lesion and non-lesion patches assigned to each cluster out of the total number of lesion and non-lesion patches of each medical center. The values are based on a random sample of 20\% of the patches in the support set of each medical center. Underlines in the table denote clusters containing the majority of lesion patches.

\begin{table}
    \caption{Sections $r_{pos}$ and $r_{neg}$ show the percentage of lesion and non-lesion patches assigned to each GMM $g$. Section $\pi_l(c, g)$ is the estimated probability (expressed as a percentage) that a patch of the cluster $g$ has of being a lesion patch (see Eq. \ref{eq:prevalence}).}
    \label{tab:gmm_results}
    \setlength{\tabcolsep}{3pt}
    \begin{tabular}{c|c|c|c|c|c|c}
    \hline
    & Cluster & Center & Center & Center & Center & Center \\
    Metric & $g$ & 0 & 1 & 2 & 3 & 4 \\
    \hline
        Lesion \% & 0 & 17.3 & 0.8 & 0.0 & 0.7 & 15.5 \\
        $r_{pos}$ & 1 & 0.1 & 23.8 & 79.8 & 0.8 & 0.1 \\
        & 2 & 7.1 & 0.5 & 13.0 & 2.9 & 0.3 \\
        & 3 & 0.0 & 0.6 & 4.4 & 78.1 & 80.1 \\
        & 4 & 0.9 & 74.2 & 0.2 & 0.2 & 0.0 \\
        & 5 & 74.6 & 0.0 & 2.7 & 11.4 & 0.1 \\
        \hline
        Non-Lesion \% & 0 & 14.6 & 14.3 & 59.0 & 10.4 & 6.5 \\
        $r_{neg}$ & 1 & 31.3 & 36.8 & 0.8 & 16.6 & 16.2 \\
        & 2 & 16.5 & 0.8 & 8.3 & 18.5 & 12.1 \\
        & 3 & 2.6 & 31.8 & 9.8 & 2.3 & 3.5 \\
        & 4 & 24.8 & 15.1 & 19.6 & 35.8 & 51.4 \\
        & 5 & 10.3 & 1.1 & 2.5 & 16.3 & 10.2 \\
        \hline
        $\pi_l(c, g)$ & 0 & 51.7 & 4.8 & 0.0 & 35.2 & 65.7 \\
        & 1 & 0.3 & 38.3 & \underline{97.1} & 4.3 & 2.8 \\
        & 2 & 28.2 & 16.7 & 56.8 & 12.6 & 19.1 \\
        & 3 & 0.0 & 1.9 & 28.1 & \underline{95.3} & \underline{94.1} \\
        & 4 & 3.2 & \underline{81.6} & 0.8 & 0.7 & 0.0 \\
        & 5 & \underline{86.3} & 0.0 & 39.7 & 38.8 & 0.0 \\
        \hline
    \end{tabular}
\end{table}
The values in the last section of Table \ref{tab:gmm_results} are computed with:
\begin{equation}
\label{eq:prevalence}
    \pi_l(c, g) := r_{pos}(c, g) \left [  r_{pos}(c, g) + r_{neg}(c, g) \right ] ^{-1}
\end{equation}
where $r_{pos}(c, g)$ is the ratio of lesion patches of medical center $c$ associated with cluster $g$, and $r_{neg}(c, g)$ is the same ratio for non-lesion patches.
The value $\pi_l(c, g)$ is the estimated density of lesion patches inside each cluster. We used this lesion probability estimate to choose the support patches to fed to the Conditioning Branch for each query patch as described in Subsection \ref{ssec:training}.

\subsubsection{Extraction of prototypes support patches}
\label{ssec:prototypes}
for each medical center, prototype patches were chosen for each GMM.
Prototypes are extracted, for each GMM, with k-means clustering on all the support patches assigned to each GMM. The number of clusters for the k-means algorithm is the number of support patches assigned to a GMM divided by a constant, called \emph{microcluster dimension}, which is set as a hyperparameter during training.
We extracted separately prototypes of patches classified as lesion and non-lesion, as such, for each GMM, two \emph{pools of prototypes} are available: the pool of lesion prototypes, which could be empty if the GMM does not contain lesion patches; the pool of non-lesion prototypes, which could be empty, when the GMM contains lesion patches only.

\subsection{Training procedure}
\label{ssec:training}

During training, the Segmentation Branch is to be fed in input by pairs $(I_q^c, L_q^c)$ of query patches and the corresponding annotation masks. $I_q^c$ is a generic query patch from center $c$ and $L_q^c$ is its annotation mask. 
In our case $c \in \{0, 1, 2\}$.
For each query patch, the Conditioning Branch expects to receive an input of $k$ patches (\emph{k-shots}) chosen among the support prototypes from the same medical center and same GMM of the query patch.
The prototypes are the ones described in Subsection \ref{ssec:prototypes} and were selected with the following policy:
\begin{enumerate}
    \item each query patch was associated with its corresponding GMM $g$;
    \item  the type and ordering of the $k$ support patches was determined as a function of the GMM $\pi_l(c, g)$ thus computed, in particular:
    \begin{itemize}
        \item the lesion probability estimate $\pi_l(c, g)$ was multiplied by $2^k$ and the result is represented with $k$ binary digits. From each binary digit of the representation we extracted the class the support prototype must belong to. If the digit is 0 the prototype was extracted from the pool of non-lesion prototypes, otherwise the prototype is extracted the from the pool of lesion prototypes (see Subsection \ref{ssec:clustering} for the procedure to create
        the pools);
        \item once the pool of prototypes was chosen, the closest support prototype, according to the L2-norm in the PCA latent space, was selected as support shot;
        \item the process continued until we completed all support shots\footnote{At each iteration, if the prototype had already been used in a previous iteration, the next closest prototype is used as support shot.};
    \end{itemize}
    \item the chosen support prototypes were concatenated together and fed to the Conditioning Branch, such that its tensor input had shape $3 \cdot k \times 128 \times 128$.
\end{enumerate}

\subsection{Training loss and regularizer}
\label{ssec:loss}

For training we used an Adam optimizer with initial learning rate $0.001$, $\beta_1$ 0.9, and $\beta_2$ 0.999. The chosen loss function was a weighted binary cross-entropy (BCE) summed with an auxiliary pretext loss \cite{Jing2019}. Formally, given a query pair $(I_q, L_q)$, we selected $k$ support shots $\{ I_s^0, \dots, I_s^{k-1} \}$ and one estimated lesion probability $\pi_l(I_q)$\footnote{For every $I_q$ we have a uniquely identified center $c$ and cluster $g$, therefore we can write $\pi_l(I_q)$ as a shorthand for $\pi_l(\mathbf{c}(I_q), \mathbf{g}(I_q))$ where $\mathbf{c}(\dot)$ and $\mathbf{g}(\dot)$ are maps from $I_q$ to its corresponding center $c$ and cluster $g$.}.
The resulting training loss was then defined as:
\begin{multline}
\label{eq:loss}
    \mathcal{L}_{train}(I_q) := \mathcal{L}_{\texttt{wBCE}} \left (M_s(I_q), L_q , w_l \right ) + \\
    w \cdot \mathcal{L}_{\texttt{BCE}} \left (\sigma \left (k^{-1}\sum_{j=0}^{k-1} M_c(I_s^j) \right ), \pi_l(I_q) \right )
\end{multline}
The first term of the sum is the weighted BCE loss between the predicted mask $M_s(I_q)$ of the query patch $I_q$ and its ground truth mask $L_q$.
The second term is the auxiliary pretext loss: namely, the BCE loss for the pretext task of estimating the density of lesion patches $\pi_l(I_q)$ from the output of a sigma function, $\sigma$, applied to the average activation of the sum of the masks $M_c(I_s^j)$ predicted by the Conditioning Branch for all $k$ shots in support set $j$. The hyperparameter weights $w_l$ and $w$ were set during training. We used early stopping to avoid overfitting: the training stopped if the validation loss did not improve for 3 consecutive epochs.

We trained our co-FCN architecture on slides from medical centers 0, 1 and 2 and used WSI of patient 75 of medical center 3 to choose the best microcluster dimension.
The code was implemented in PyTorch 1.6 \cite{Paszke2019} with the use of the PyTorch Lightning 0.9 framework \cite{falcon2019pytorch}. Training took a few hours on a workstation equipped with a NVIDIA RTX\textsuperscript{\texttrademark} 2080 Ti GPU.
During inference, the same policy was applied to associate the patches of any input WSI to their corresponding support patches.

\section{Results}
\label{sec:results}

For testing we used six WSIs from medical centers 3 and 4 of the CAMELYON17 dataset, three WSIs for each medical center (see Section \ref{ssec:dataset} for details). The WSIs were selected not to be part of the support set of centers 3 and 4 and such that the annotated slides contained all three different types of metastases in the dataset.

We classified a patch as lesion if at least one pixel in its central area, $64 \times 64$ pixels wide, was annotated as lesion.
For each patch, the predicted probability that the patch had of being a lesion patch (lesion probability, for short) was the minimum over the probabilities of any of its pixels in its central region had of being classified as lesion.

In the following sections we identify each WSI with the notation `patient ID/node ID', where the `patient ID' is a unique patient identifier in the CAMELYON17 dataset, and the `node ID' is a unique identifier of the slides collected for each patient.

\subsection{Benchmark}

We compared the co-FCN against a baseline U-Net architecture, configured as the Segmentation Branch.
The U-Net was trained with the same Adam optimizer, BCE loss, and hyperparameters of the co-FCN.
The dataset used for training of the U-Net was the union of the support and query sets for centers 0, 1 and 2. Adding the support set of these three centers to the training set of the U-Net increased the dataset size by approximately 40\%. 

To compare the performance of our co-FCN against the U-Net
we used the Area Under the Curve (AUC) of the Receiver Operating Characteristic (ROC), computed with the \texttt{pROC} R package \cite{Robin2011}, on each separate WSI.

Table \ref{tab:auc-unet} shows the results obtained by performing inferences on the WSIs of medical centers 3 and 4 by the U-Net. Along with the AUC mean values we also include their 95\% confidence interval (CI), computed with the \texttt{ci.auc} function of the \texttt{pROC} package: this function applies the method described in \cite{DeLong1988} and implemented by \cite{Sun2014}. The CI is computed with the R quantile function \texttt{qnorm}.

Higher than 0.97 AUC scores were obtained on all WSIs except for the slide with micro-metastases of patient 67 (center 3) and a slide containing ITCs of patient 89 (center 4).
\begin{table}
\caption{AUC with 95\% CI of U-Net for center 3 and 4 WSIs.}
\label{tab:auc-unet}
    \setlength{\tabcolsep}{3pt}
    \centering
\begin{tabular}{lllll}\hline
\textbf{Center ID}&\textbf{ITC} &\textbf{micro} &\textbf{macro} \\\hline
3 &$0.981 \pm 0.011$ &$0.533 \pm 0.077$ &$0.982 \pm 0.003$ \\
4 &$0.577 \pm 0.068$ &$0.943 \pm 0.008$ &$0.989 \pm 0.002$ \\
\hline
\end{tabular}
\end{table}

\subsection{Comparison}

In Tables \ref{tab:fsl-ccat-none-perc-c3} and \ref{tab:fsl-ccat-none-perc-c4} we show the percentage difference among the AUCs of the two correlated ROC curves, one from the prediction of the co-FCN and the other from the prediction of the corresponding U-Net, with the \texttt{roc.test} function of the \texttt{pROC} R package. This function tests (see \cite{DeLong1988}) if AUCs showed a statistically significant difference\footnote{For each two AUCs comparison, the test outputs a $\textit{p-value}$ which we record in the tables next to the percentage difference using the following significance codes:
\begin{itemize}
    \item *** - $0 \leq \textit{p-value} < 0.001$
    \item ** - $0.001 \leq \textit{p-value} < 0.01$
    \item * - $0.01 \leq \textit{p-value} < 0.05$
    \item . - $0.05 \leq \textit{p-value} < 0.1$
\end{itemize}
For $\textit{p-value}$ larger than 0.1 only the percentage difference is shown.}.

For the co-FCN, we tested with different number of shots: 1, 2, 4 and 8 shots.
All training experiments were conducted with the same settings: initial learning rate 0.001; lesion class weight ($w_l$ in Eq. \ref{eq:loss}) 4.0; pretext task loss weight ($w$ in Eq. \ref{eq:loss}) 0.2; microcluster dimension 20 (see Subsection \ref{ssec:prototypes}); 75\% of the patches in the query set were used as training and the rest are left for validation; we used early stopping so that the training was stopped if there were no improvement in the validation loss for 3 consecutive epochs.

As seen in Tables \ref{tab:fsl-ccat-none-perc-c3} and \ref{tab:fsl-ccat-none-perc-c4}, a statistically significant improvement in the AUC score was obtained, for all shots greater than 1, for the WSI with micro-metastases of center 4 (patient 88). A marginal improvement can also be seen on the WSI with ITCs (patient 89) of center 4.
On all other WSIs the conditioning did not improve the AUC scores and for both WSIs with macro-metastases, patient 75 of medical center 3 and patient 99 of medical center 4, it resulted in a marginal degradation of the AUC performance.
Overall, the best performances was achieved with 8 shots.
\begin{table}
    \caption{Percentage difference of mean AUC of co-FCN w.r.t. baseline U-Net for Center 3 WSIs.}
    \label{tab:fsl-ccat-none-perc-c3}
    \setlength{\tabcolsep}{3pt}
    \centering
    \begin{tabular}{p{44pt}|p{44pt}|p{44pt}|p{44pt}}
    \hline
    &72/0 &67/4 &75/4 \\
    Shots &ITC &micro &macro \\
    \hline
    1 &-0.9\% &-5.1\% &-1.2\% ***\\
    2 &-2.5\% ** &  1.9\% &-5.6\% ***\\
    4 &-1.8\% &22.1\% . &-1.4\% ***\\
    8 &-1.0\% &  4.1\% &-1.9\% ***\\
    \hline
\end{tabular}
\end{table}
\begin{table}
    \caption{Percentage difference of mean AUC of co-FCN w.r.t. baseline U-Net for Center 4 WSIs.}
    \label{tab:fsl-ccat-none-perc-c4}
    \setlength{\tabcolsep}{3pt}
    \centering
    \begin{tabular}{p{44pt}|p{44pt}|p{44pt}|p{44pt}}
    \hline
    &89/3 &88/1 &99/4 \\
    Shots &ITC &micro &macro \\
    \hline
    1 &0.3\% &0.5\% &-0.3\% *\\
    2 &5.7\% &5.3\% *** &-1.4\% ***\\
    4 &14.7\% . &4.8\% *** & 0.0\%\\
    8 &22.7\% * &5.0\% *** &-0.3\% *\\
    \hline
\end{tabular}
\end{table}

Other minor architectural variants were tested. In particular, we tried different connections between the two branches. The results were in line with what already discussed for the `features concatenation' and confirmed that the co-FCN is more effective than the U-Net for WSIs with ITC and micro-metastases, from medical center 4.

The same behaviour was confirmed in an experiment where, for training both for the co-FCN and for the U-Net, we replaced the WSIs of medical center 4 with the WSIs of medical center 2.
The prediction on two WSIs from medical center 2 were then compared between the U-net and the co-FCN (for the co-FCN we used microcluster dimension 20 and 8 shots).
In this other experiment we observed again that the AUCs of the co-FCN were higher than the AUCs of the U-net for the WSIs with ITCs and micro-metastasis: the co-FCN AUC for the WSI with ITCs improved 16\% over the U-Net (from 0.57 to 0.66) and the co-FCN AUC for the WSI with micro-metastases improved 7\% (from 0.90 for the U-Net to 0.96 for the co-FCN).

\subsection{Visual comparison of co-FCN and U-Net predictions}

\begin{figure}
    \begin{subfigure}{0.45\textwidth}
        \includegraphics[width=\textwidth]{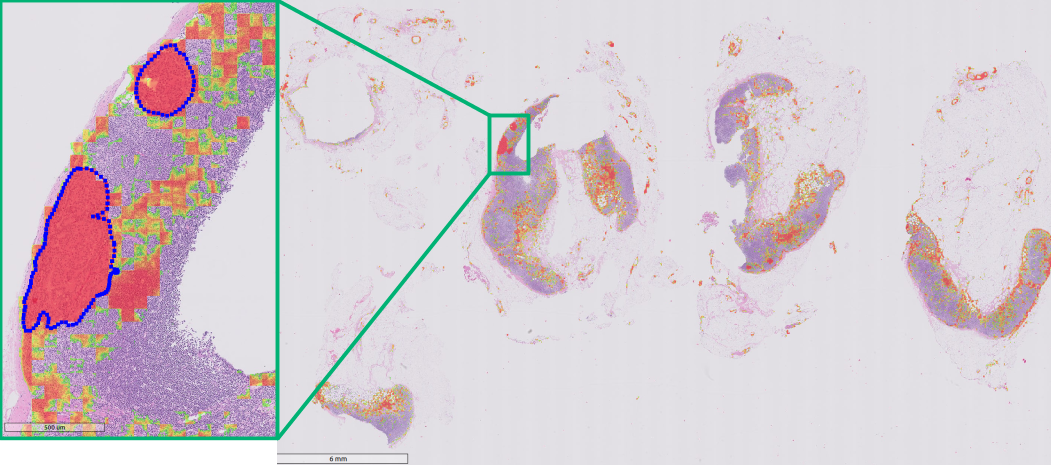}
        \caption{co-FCN}
        \label{fig:p88_n1_ccat_none_s8}
    \end{subfigure}
    \begin{subfigure}{0.45\textwidth}
        \includegraphics[width=\textwidth]{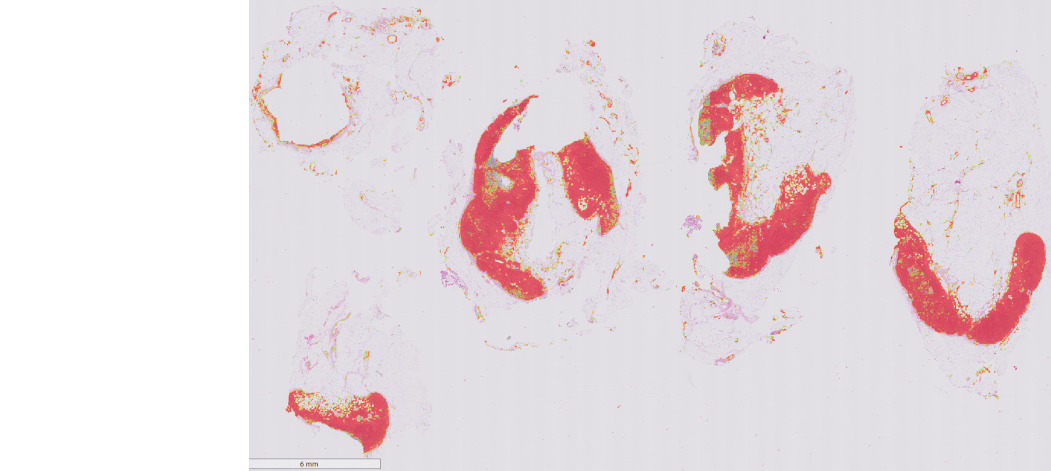}
        \caption{U-Net}
        \label{fig:p88_n1_unet}
    \end{subfigure}
    \caption{Micro-metastases in patient 88 of medical center 4 as detected by the co-FCN at 8 shots (a) and by the U-Net (b). Prediction probabilities below 0.75 are transparent, between 0.75 and 1.0 are shown with hues changing from green to red.}
    \label{fig:pred_p88_n1}
\end{figure}

Apart from the AUCs comparison, it is instructive to compare the prediction results of the co-FCN and of the U-Net on the two WSIs of medical center 4 with ITCs and micro-metastases.
The results can best be seen in Fig. \ref{fig:p88_n1_ccat_none_s8}.
The comparison with the predictions from the regular U-Net (see Fig. \ref{fig:p88_n1_unet}) highlights the lower specificity of the U-Net.

\begin{figure}
    \begin{subfigure}{0.45\textwidth}
    \includegraphics[width=\textwidth]{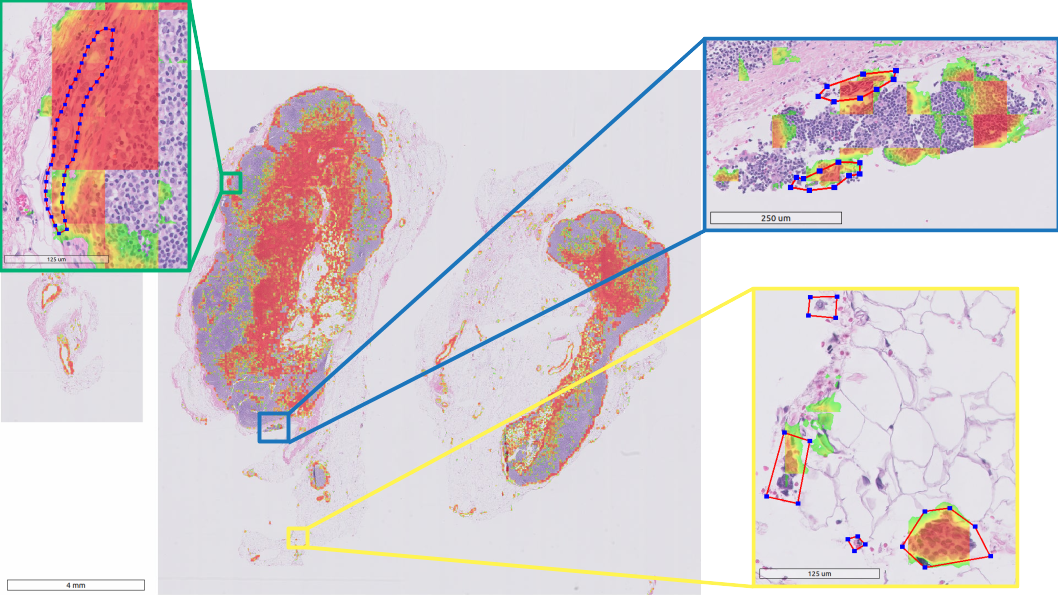}
        \caption{co-FCN}
        \label{fig:p89_n3_ccat_none_s8}
    \end{subfigure}
    \begin{subfigure}{0.45\textwidth}
        \includegraphics[width=\textwidth]{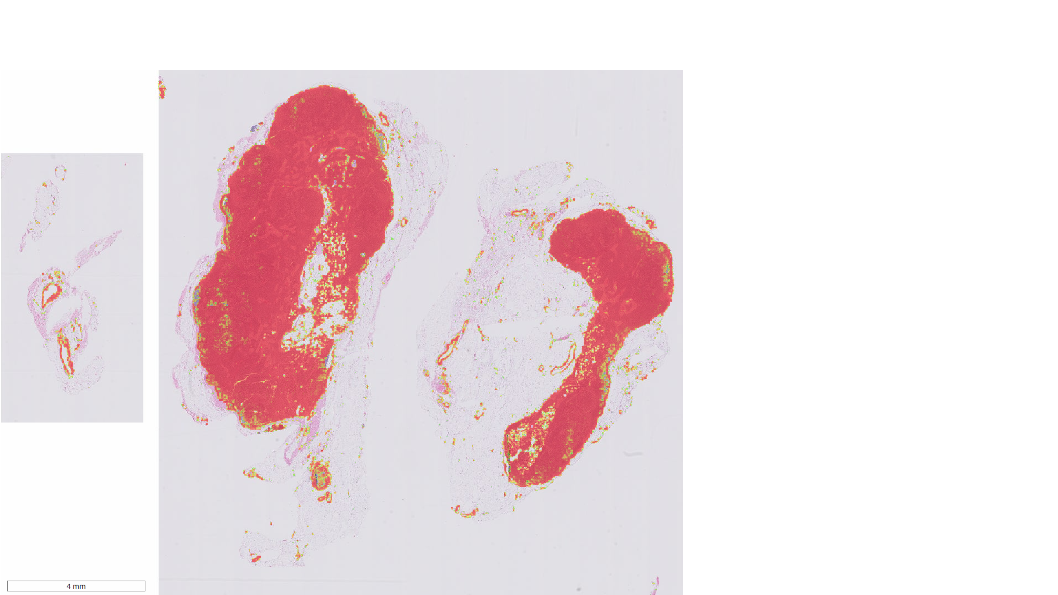}
        \caption{U-Net}
        \label{fig:p89_n3_unet}
    \end{subfigure}
    \caption{ITCs in patient 89 of medical center 4 as detected by the co-FCN at 8 shots (a) and by the U-Net (b). Prediction probabilities below 0.75 are transparent, the others are shown with hues from green (0.75) to red (1.0).}
    \label{fig:pred_p89_n3}
\end{figure}

The same behavior can be seen on the WSI with ITCs from medical center 4 (patient 89, node 3).
The different behaviour is evident in Fig. \ref{fig:pred_p89_n3} where the co-FCN heatmap (Fig. \ref{fig:p89_n3_ccat_none_s8}) is compared against the U-Net heatmap (Fig. \ref{fig:p89_n3_unet}). The co-FCN detects all except one of the ITCs, but the U-Net classifies all the regions with tissue as potential lesion. The co-FCN is also conservative, but it does a better job in discriminating areas where ITCs could be present from the rest.

\section{Discussion and conclusions}

We introduced a co-FCN architecture, derived by \cite{GuhaRoy2020}, and we adapted this architecture to the histopathology domain.
For the co-FCN to work well with histopathology slides we learnt that the selection of the conditioning images (patches extracted from the support WSIs) is critical for the co-FCN to outperform a U-Net architecture on the same segmentation task. Therefore we developed an automated selection protocol for the conditioning images which we summarize below.
Our results show that the co-FCN achieves better performance over a similarly trained U-Net, in that it predicts lesion regions on WSIs, containing ITCs and micro-metastases, collected with a digital scanner and from a medical center not used for the collection of the training WSIs.

\subsection{Domain adaptation}

In reference to WSIs from medical center 4, which uses a digital slide scanner not used in any other centers, the benchmark U-Net contains more false positives than the co-FCN trained on the same set of WSIs but conditioned on a support set from center 4. This is true in particular for the more difficult WSIs containing ITCs and micro-metastases:
\begin{itemize}
    \item the best AUC for the WSI of patient 89 (containing ITCs) obtained by the U-Net is $0.589 \pm 0.070$, a result which is outperformed by 20\% with a co-FCN at 8 shots: $0.708 \pm 0.091$;
    \item the best AUC for the WSI of patient 88 (containing micro-metastases) obtained by the U-Net $0.943 \pm 0.008$ is slightly improved by 5\% with a co-FCN at 8 shots: $0.990 \pm 0.006$
\end{itemize}
The difference in the latter case, due to the high number of false positives raised by the U-Net, is more evident when applying the pAUC metric\footnote{The \emph{partial} AUC (pAUC) is a measure of the AUC of the ROC curve over a range of interest, either a specificity or sensitivity range \cite{Walter2005}.} in the specificity range 90\%-100\%. In this case the U-Net tops at 0.720 pAUC a result improved by 34.4\% by the co-FCN, which achieves a pAUC of 0.968.

The co-FCN therefore performs better than a standard U-Net when facing domain shift and could be used, without retraining, to screen WSIs for the possible presence of ITCs and micro-metastases.

\subsection{Advantages of our support set selection method}

Finally, we note that an advantage of our support selection method is that, to compute the lesion probability estimate $\pi_l$ for each cluster (see Subsection \ref{ssec:clustering}), only patch-level classification is required, instead of a full dense annotation of the example WSIs from where the patches are extracted.
This allows for the usage of sparse annotations to create the reference set for selecting support shots, 
with a net reduction in the annotation effort by pathologists.
This is in keeping with the findings by Rakelly et al. \cite{Rakelly2018} in their architecture who also used sparse annotations in their support set.

Future work will explore the interaction between the latent representation extracted by the Conditioning Branch and the Segmenter output. Furthermore, as the selection of the support set is a crucial step for the successful training and inference of the co-FCN, we believe that the study of unsupervised strategies and methods for the proper selection of the support set remains a promising research topic.

\addtolength{\textheight}{-6cm}   

\end{document}